\documentclass[%
superscriptaddress,
 amsmath,amssymb,
 aps, physrev,
 twocolumn,
prl,
]{revtex4-2}

\usepackage{color}
\usepackage{graphicx}
\usepackage{dcolumn}
\usepackage{bm}
\usepackage[whole]{bxcjkjatype} 
  \usepackage{hyperref}
  \hypersetup{pdfpagemode={UseOutlines},
  bookmarksopen=true,
  bookmarksopenlevel=0,
  hypertexnames=false,
  colorlinks=true,
  citecolor=blue,
  linkcolor=blue,
  urlcolor=blue,
  allcolors=blue,
  pdfstartview={FitV},
  unicode,
  breaklinks=true,
  }

\begin{document}


\title{
Chirality in $(\vec{p},2p)$ reactions induced by proton helicity
}%

\author{Tomoatsu Edagawa}
\affiliation{
Research Center for Nuclear Physics, Osaka University, Ibaraki 567-0047, Japan
}

\author{Kazuki Yoshida}
\email{Contact author: yoshidak@rcnp.osaka-u.ac.jp}
\affiliation{
Research Center for Nuclear Physics, Osaka University, Ibaraki 567-0047, Japan
}
\affiliation{ 
Interdisciplinary Theoretical and Mathematical Sciences Program (iTHEMS), RIKEN, Wako 351-0198, Japan
}

\author{Shoichiro Kawase}
\affiliation{
Department of Advanced Energy Science and Engineering, Kyushu University, Kasuga, Fukuoka 816-8580, Japan
}
\author{Kazuyuki Ogata}
\affiliation{
Department of Physics, Kyushu University, Fukuoka 819-0395, Japan
}
\affiliation{
Research Center for Nuclear Physics, Osaka University, Ibaraki 567-0047, Japan
}

\author{Masaki Sasano}
\affiliation{
RIKEN Nishina Center, Hirosawa 2-1, Wako, Saitama 351-0198, Japan
}

\date{\today}

\begin{abstract}
  It is shown that longitudinally polarized protons can be used to induce chirality in the final states of the $(\vec{p},pN)$ reaction at intermediate energies, when there exist three final-state particles with non-coplanar momentum vectors. The analyzing power $A_z$ is proposed as a measure of this effect.
Theoretical descriptions to obtain $A_z$ based on an intuitive picture as well as a distorted wave impulse approximation are presented, showing that the helicity of incident protons is coupled to the chirality of the orbital motion of a single-particle wave function, resulting in the chirality of the final states and a large $A_z$ value.
\end{abstract}

\maketitle

\textit{Introduction.}
The origins of chirality in a variety of materials such as neutrinos, optical isomers, and biological molecules have attracted a wide range of interests~\cite{GOLDHABER1958,HEGSTROM1982,MASON1984}.
In nuclei, there are theoretical interpretations to explain doublet bands observed for triaxial nuclei as evidence of the spontaneous chiral symmetry breaking of nuclear structures in the intrinsic frame~\cite{FRAUENDORF1997,HAMAMOTO2013}.
However, in the laboratory frame, the total Hamiltonian of an isolated nucleus is invariant under mirror reflection.
Therefore, the probabilities of two mirror-symmetric nuclear phenomena are the same (thus, there is no chirality), unless an external stimulus that breaks the mirror-reflection symmetry is introduced to the system. 
A typical example of such a stimulus is the longitudinally spin-polarized beam, as considered below.
The exceptions are weak processes, e.g., the parity violation in the $\beta$ decay, which can be chiral without the help of an external stimulus violating the mirror-reflection symmetry. 
Also, in the neutron-skin thickness determination through parity-violating electron scattering in the PREX/CREX experiments~\cite{PREX2021,CREX2022}, the weak interaction ($Z^0$ boson exchange) acts as ``an external stimulus'' that violates the mirror-reflection symmetry.
In this case, the induced chirality is observed as the dependence of the cross section on the incident-electron helicity.

There are also many studies with transversely polarized beams for low/intermediate-energy nuclear reactions~\cite{HAEBERLI1967,GLASHAUSSER1987,Tamii2011,SEKIGUCHI2002} and high-energy hadron production~\cite{Adams1991_1,Adams1991_2,Adams2004}. 
Previous studies of knockout reactions using transversely polarized beams are discussed below.
Compared to those works, a unique point of the present work is that the initial state is characterized by the helicity (longitudinal polarization). 
We note that $A_y$ ($A_N$ in high-energy hadron production) can be theoretically related to helicity-dependent transition matrices because a given helicity state can be written as a linear combination of transverse spin states.

In this Letter, we propose to use a longitudinally spin-polarized proton beam as a source of chirality in the $(\vec{p},pN)$ reaction at intermediate energies and investigate how the effects associated with chirality appear in the reaction.
This type of reaction has been widely used to investigate single-particle (s.p.) structure of nuclei~\cite{JACOB_MARIS1966,JACOB_MARIS1973,WAKASA2017,NORO2020}: 
The nucleon knockout reaction can be regarded as a quasifree nucleon-nucleon ($NN$) scattering of the incident proton and a nucleon inside the target nucleus.
From the kinematics of the particles in the final state, i.e., the recoil proton, the knocked-out nucleon, and the reaction residue, the s.p.~properties of the struck nucleon, e.g., the orbital angular momentum, the s.p.~energy, and the spectroscopic factor, can be reliably obtained.

To the best of our knowledge, in past studies using the $(\vec{p},pN)$ reaction on nucleus targets, only the transverse polarization of incident protons is used to measure the so-called vector analyzing power $A_y$~\cite{JACOB_MARIS1966,JACOB_MARIS1973,WAKASA2017,NORO2020,NORO2023}, which appears as the geometrical asymmetry between the left and right scatterings on the scattering plane in the coplanar condition. The to-be-knocked-out nucleon in the initial state is effectively polarized orthogonal to the scattering plane, because of the Maris effect~\cite{JACOB_MARIS1973,MARIS1980,SHUBHCHINTAK2018}.
The Maris effect originates from two characteristics of the $(\vec{p},2p)$ reaction.
One is the large spin correlation coefficient $C_{yy}$ and the cross section for a spin-parallel pair is larger than those for an antiparallel pair~\cite{NNONLINE}.
The other is the absorption effect by the $p$-nucleus optical potential.
In consequence, the Maris effect gives large $A_y$ and characteristic emission energy dependence, as typically shown in Ref.~\cite{Kitching80}.

Herein, we polarize the incident proton along the beam direction and measure the asymmetry appearing as chirality defined by the momentum vectors (collectively labeled as $\mathcal{K}$) for the three particles of the final states. The knocked-out nucleon is effectively polarized in parallel to the beam direction.

To evaluate this effect, we introduce the longitudinal vector analyzing power $A_z$ as
\begin{equation}
  A_z 
  = 
  p_z \frac{\sigma_+(\mathcal{K}) - \sigma_+(\mathcal{\tilde{K}})}{\sigma_+(\mathcal{K}) + \sigma_+(\mathcal{\tilde{K}})},
  \label{eq:Az}
\end{equation}
where $p_z$ is the polarization of the beam particle in parallel with the beam direction and $\sigma_{\pm}(\mathcal{K})$, where the subscript of $\sigma$ denotes the polarization direction, are the cross sections associated with $\mathcal{K}$ and its mirror partner $\mathcal{\tilde{K}}$, respectively.
For simplicity, we assume a pure state of the polarization $(p_z = \pm1)$ hereafter.
One can easily find an equivalent definition as 
\begin{equation}
    A_z 
    = 
    \frac{\sigma_{+}(\mathcal{K}) - \sigma_{-}(\mathcal{K})}{\sigma_{+}(\mathcal{K}) + \sigma_{-}(\mathcal{K})},
    \label{eq:Az2}
\end{equation}
which is defined as the asymmetry between $\sigma_+$ and $\sigma_-$ for one of the mirror partners $\mathcal{K}$.

Apart from the $(p,pN)$ reaction, $A_z$ was measured for several different probes in the past: 
Recently, the electron-helicity-dependent cross section and its asymmetry have been studied in quasielastic $^{40}\mathrm{Ca}(e,e'p)^{39}\mathrm{K}$ reaction from $1d_{3/2}$ orbital in the non-coplanar condition~\cite{Kolar25}.
Although the asymmetry is discussed in a different reaction and context, the asymmetry reported in Ref.~\cite{Kolar25} should be related to the present study.
Also, in $\vec{d}\vec{p}$ breakup reaction in the non-coplanar condition~\cite{Meyer04}, $A_z$ has been measured and analyzed using CD-Bonn~\cite{Machleidt01} and AV18~\cite{Wiringa95} $NN$ interactions in a Faddeev calculation~\cite{Glockle96}.
We note that, in these existing studies, the obtained values of $A_z$ are not so large, because the asymmetry arises from effects other than the Maris effect.

\begin{figure}[tbp]
  \centering
  \includegraphics[width=\linewidth]{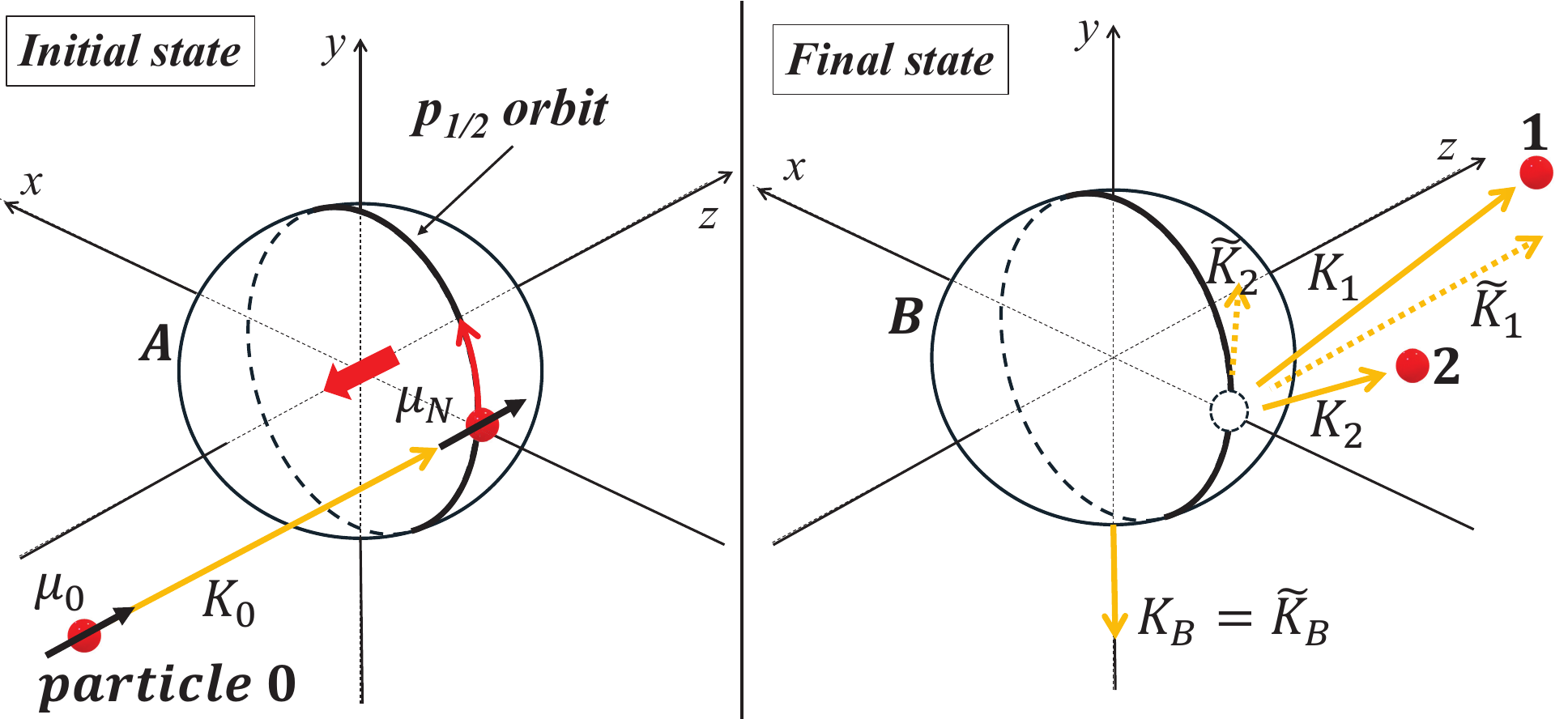}
  \caption{
      A schematic view of initial (left) and final (right) states in the $^{16}$O$(\vec{p},pN)$ reaction, where $0p_{1/2}$ proton having the spin (orbital angular momentum) in parallel with $z$ ($-z$) direction is knocked out.
      As $C_{zz} \approx 1$, the $\mu_0 = \mu_N$ case is considered.
      Note that $\mathcal{K}$ must be non-coplanar so as to satisfy $\mathcal{K}\neq\mathcal{\tilde{K}}$.
      Here we consider the laboratory frame where the nucleus $A$ is at rest.
}
    \label{fig:reaction}
\end{figure}

This Letter is structured as follows.
First we show the origin of the asymmetry in $A_z$ by providing an intuitive explanation, which is followed by a theoretical basis as well as a realistic calculation result.  
As a simple case, we study the $^{16}$O$(\vec{p},2p)^{15}$N reaction at the incident proton energy of 250~MeV, where two different orbits, $0p_{1/2}$ and $0p_{3/2}$, are considered to evaluate $A_z$, followed by discussions and conclusions.

\begin{figure}[tbp]
    \centering
    \includegraphics[width=\linewidth]{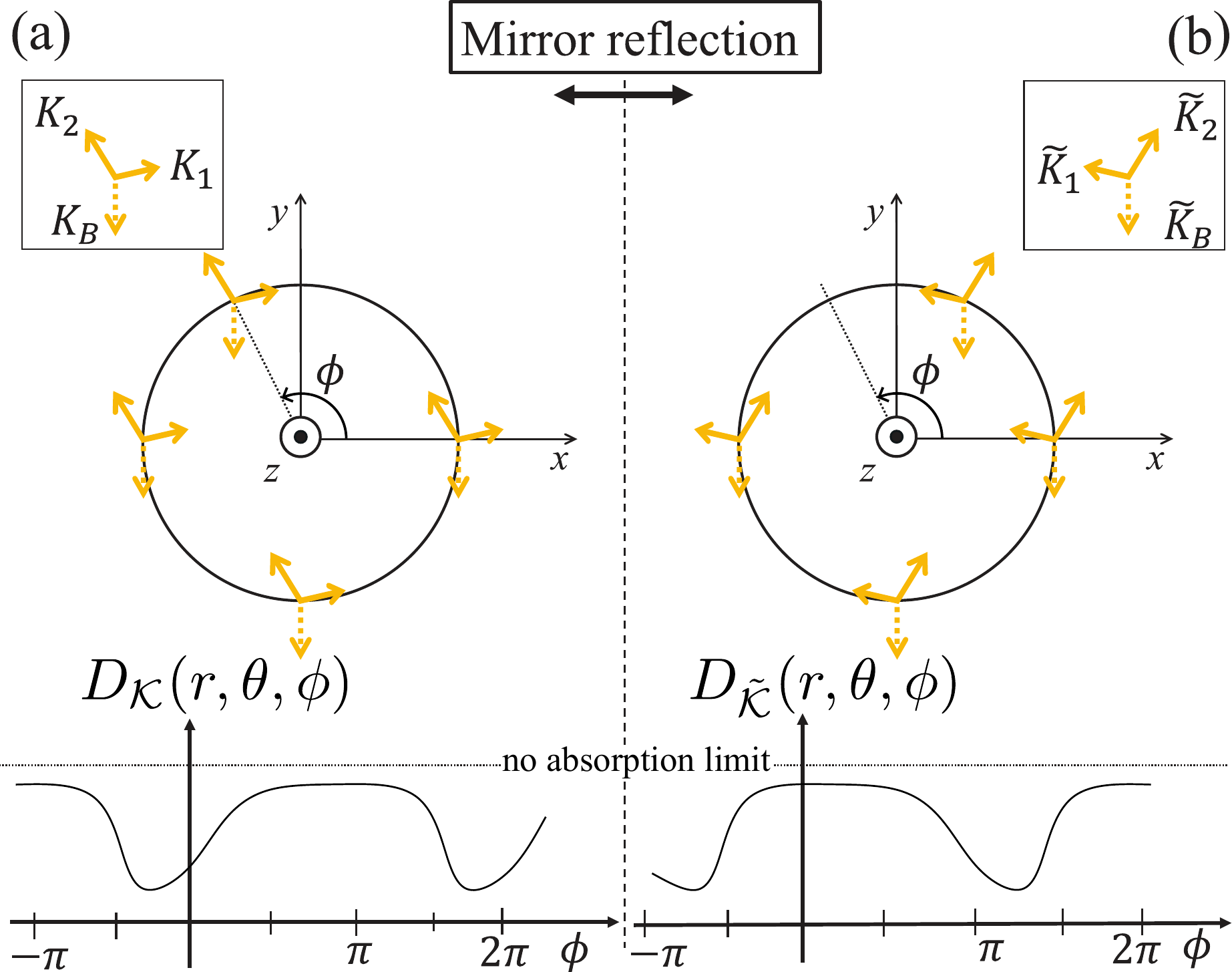}
    \caption{
    Left panel (a): (Inset on top) $\mathcal{K}=(\bm{K}_1,\bm{K}_2,\bm{K}_\mathrm{B})$ in the final state on the $xy$-plane. 
    Note that only the transverse component of $\mathcal{K}$ is shown in the figure, and the longitudinal component of $\bm{K}_2$ is smaller than $\bm{K}_1$. 
    (Middle) A geometrical relation between $\mathcal{K}$ and $\phi$. 
    The absorption is maximum in $3\pi/2 \lesssim \phi \lesssim 2\pi$, where particle 2 ($\bm{K}_2$) goes across the nucleus in this example.
    (Bottom) A schematic figure of the $\phi$ dependence of $D_{\mathcal{K}}$ as a function of $\phi$ for a certain $(r,\theta)$. 
    Right panel (b): Same as (a) but for the mirror partner $\mathcal{\tilde{K}}$.
    Note that $D_{\mathcal{K}}$ and $D_{\mathcal{\tilde{K}}}$ are both periodic with respect to $\phi$ and related as $D_{\mathcal{K}}(r,\theta,\phi) = D_{\mathcal{\tilde{K}}}(r,\theta,\pi-\phi)$. 
    }
    \label{fig:Dk}
\end{figure}

\textit{Formalism.}
In the present paper, the target and residual nuclei are labeled as A and B, respectively, and the initial proton, emitted proton, knocked-out proton and the proton in A to be struck are labeled as particles 0, 1, 2, and $N$, respectively.
The momentum (wave number) of particle/nucleus $i$ ($i$ = 0, N, 1, 2, A, or B) is denoted by $\bm{K}_i$, and the third component of the spin of the particle $i$ is denoted by $\mu_i$.
A set of the momenta in the final state is denoted as $\mathcal{K}=(\bm{K}_1,\bm{K}_2,\bm{K}_\mathrm{B})$ and its mirror reflection (i.e. $x \rightarrow -x$) is denoted by $\mathcal{\tilde{K}}=(\tilde{\bm{K}}_1,\tilde{\bm{K}}_2,\tilde{\bm{K}}_{\mathrm{B}})$.
The nucleon momentum inside the target nucleus in the initial state is denoted by $\bm{K}_N$.
In the quasifree $(p,pN)$ reaction, we chose kinematical conditions that the $p$-$N$ elementary process approximately satisfies the $p$-$N$ elastic scattering kinematics,  $\bm{K}_0 + \bm{K}_N \approx \bm{K}_1 + \bm{K}_2$ and $\bm{K}_\mathrm{B} \approx -\bm{K}_N$.
Figure~\ref{fig:reaction} shows a schematic view of initial and final states of the $^{16}$O$(\vec{p},pN)$ reaction from $0p_{1/2}$ orbit with $\mu_0=\mu_N$.
There are three steps in our intuitive explanation: 
Firstly, in this beam-energy domain, the $NN$ scattering amplitude has a strong spin dependence;
Cross sections for a spin-parallel pair of the colliding nucleons are much higher than those for an antiparallel pair ($C_{zz} \approx 1$) for a broad range of scattering angles~\cite{NNONLINE}.
Consequently, the helicity of the incident proton ($\mu_0$) is coupled to that of the s.p. orbit ($\mu_N$) with the same direction, $\mu_0=\mu_N$.
The spin direction of $\mu_0$, $\mu_N$, and the orbital angular momentum of the proton $\bm{l}$ are, $\mu_0 \parallel \mu_N \parallel \hat{\bm{z}}$ and $\bm{l} \parallel -\hat{\bm{z}}$, in this case.
This orbital motion, which is transverse to the beam direction, makes the $NN$-scattering plane tilted with respect to the beam axis.
As a result, $\mathcal{K}$ and $\mathcal{\tilde{K}}$ become non-coplanar. 
Secondly, the emitted particle 1 has larger momentum and kinetic energy than particle 2, which corresponds to forward scattering of the quasi-free $NN$ scattering. 
Lastly, the asymmetry between events with $\mathcal{K}$ and $\mathcal{\tilde{K}}$ appears as a result of the nuclear absorption effect.
In events with $\mathcal{K}$, the momentum of particle $1$ is large enough to go through the nucleus.
At the same time, particle $2$ with smaller momentum can also be observed without being absorbed, because particle $2$ needs to penetrate only a short distance, as shown in Fig.~\ref{fig:Dk}.
On the other hand, in $\mathcal{\tilde{K}}$, the particle $2$ is barely observed, because it has to go across the nucleus with small momentum. 
Consequently, the cross section with $\mathcal{K}$ is larger than that with $\tilde{\mathcal{K}}$, thereby resulting in a positive value of $A_z$ in Eq.~(\ref{eq:Az}).

A more realistic description is given by the distorted-wave impulse approximation (DWIA), according to Ref.~\cite{OGATA2024}. 
The DWIA calculations in the present paper are done using the transition matrix of Eq.~(1) of Ref.~\cite{OGATA2024}. 
Herein only essential parts are described below.
We note that we omit the spin rotation of the distorted waves caused by nuclear spin-orbit (LS) potential for an explanatory purpose.
In the realistic calculation shown below, these degrees of freedom are exactly treated.
We follow the Madison convention for the definition of the coordinate system and spin observable.

For a given $\mathcal{K}$, $\sigma_{\pm}$ is obtained as
\begin{equation}\label{eq:sigma}
  \sigma_{\pm}(\mathcal{K}) 
  =
  F_{\rm kin}(\mathcal{K}) 
  \sum_{\mu_1\mu_2\mu_j} 
  \left| 
    T_{\mu_1\mu_2,\pm1/2,\mu_j}(\mathcal{K}) 
  \right|^2,
\end{equation}
where $F_{\rm kin}(\mathcal{K})$ is the kinetic factor (i.e. the product of the phase-space volume and Jacobian) and $T_{\mu_1,\mu_2,\mu_0,\mu_j}$ represents the transition matrix of the $(p,pN)$ process, 
given by
\begin{align}
      T_{\mu_1\mu_2\mu_0\mu_j} 
      =  
      &\sum_{\mu_N} 
      \tilde{t}_{\bm{\kappa}'\mu_1\mu_2,\bm{\kappa}\mu_0\mu_N} \nonumber \\
      & \times \int d\bm{R} D_{\mathcal{K}}(\bm{R}) \Phi_{nlj,\mu_N}(\bm{R}),
    \label{eq:transition_matrix}
\end{align}
where $\tilde{t}$ is the $NN$ transition amplitude with the initial and final $NN$ relative momentum, $\bm{\kappa}$ and $\bm{\kappa}'$, respectively.
$\bm{R}$ represents the coordinate of the center-of-mass of particles 0 and $N$ regarding the nucleus B. 

The function $D_{\mathcal{K}}$ describes to what extent the reaction is hindered due to the absorption effect at coordinate $\bm{R}$, given by the product of the three distorted waves given as
\begin{equation}    
D_{\mathcal{K}}(\bm{R}) = \chi^{(-)*}_{1,\bm{K}_1}(\bm{R})
        \chi^{(-)*}_{2,\bm{K}_2}(\bm{R})
        \chi^{(+)}_{0,\bm{K}_0}(\bm{R}).
\end{equation}
The s.p. wave function is given by
\begin{equation}
    \Phi_{nlj,\mu_N}(\bm{R}) = \sum_m (lm, 1/2,\mu_N|j\mu_j)\Psi_{nljm}(\bm{R}),
\end{equation}
where $\Psi_{nljm}(\bm{R})$ is the s.p. wave function of the magnetic substate $m$ for given radial quantum number $n$, orbital angular momentum $l$, and total angular momentum $j$.
When $l$ is zero (i.e. $s$ orbit), $D_K$ and $D_{\tilde{K}}$ are uniformly integrated with respect to $\phi$, resulting in the same value (no chirality). 
On the other hand, if $l$ is finite, they are non-uniformly integrated because of the $\phi$ dependence in $\Psi_{nljm}$ (i.e. $e^{im\phi}$). 
Note that the factor $e^{im\phi}$ differs from its mirror image ($e^{-im\phi})$ (chiral), when $m \neq 0$.

Since the function $D_{\mathcal{K}}$ is periodic with respect to $\phi$, it can be Fourier-expanded with respect to $\phi$ in its domain $[ 0, 2\pi]$ as
\begin{equation}\label{eq:Dk}
    D_{\mathcal{K}}(r,\theta,\phi) = \sum_{m_d} d^{m_d}_{\mathcal{K}}(r,\theta)e^{-i m_d \phi}.
\end{equation}
Using this expansion, we rewrite the transition matrix element as a sum of $m$, 
\begin{align}
  T_{\mu_1\mu_2\mu_0\mu_j} 
  = 
  & \sum_{m} (lm,1/2,\mu_N|j\mu_j)
  \tilde{t}_{\bm{\kappa}'\mu_1\mu_2,\bm{\kappa}\mu_0\mu_N} \nonumber  \\
  & \times \mathcal{F}_{nljm}(\mathcal{K}),
  \label{eq:transition_matrix_expanded}
\end{align}
where 
\begin{align}
  \mathcal{F}_{nljm}(\mathcal{K}) 
  = 
  & 2\pi \int r^2 dr \int d(\cos\theta) \, 
  d^m_{\mathcal{K}}(r,\theta) \nonumber \\
  & \times u_{nlj}(r)P_{lm}(\cos\theta).  
\label{eq:Fnljm}
\end{align}
Here $u_{nlj}(r)$ is the radial part of $\Psi_{nljm}(\bm{R})$, and $P_{lm}(\theta)$ is the associated Legendre polynomial.
We note that the integration over $\phi$ is done and in consequence, only the $m_d = m$ component has a non-zero value.

It is noteworthy that $\mathcal{F}_{nljm}(\mathcal{K})$ and $d^m_{\mathcal{K}}$ are chiral:
\begin{equation}
    \mathcal{F}_{nljm}(\mathcal{K}) \neq \mathcal{F}_{nljm}(\mathcal{\tilde{K}}) 
        = \mathcal{F}_{nlj,-m}(\mathcal{K}),
\end{equation}
because
\begin{equation}
    d^m_{\mathcal{K}}(r,\theta) 
    \neq d^{m}_{\mathcal{\tilde{K}}}(r,\theta) 
    = d^{-m}_{\mathcal{K}}(r,\theta).
\end{equation}  
The relation $\mathcal{F}_{nljm}(\mathcal{K}) \neq  \mathcal{F}_{nlj,-m}(\mathcal{K})$ means the angular momentum vector $\bm{l}$ is ``polarized'', when $\mathcal{K}$ is selected, as intuitively explained.

Inserting Eq.~(\ref{eq:transition_matrix_expanded}) into Eq.~(\ref{eq:sigma}) and assuming that the $NN$ scattering occurs only for spin-aligned pairs (i.e. $\mu_0 = \mu_N$), we find
\begin{align}
  \sigma_{\pm}(\mathcal{K}) 
  = 
  & F_{\rm kin}(\mathcal{K})
  \left(\sum_{\mu_1\mu_2}\left| \tilde{t}_{\bm{\kappa}'\mu_1\mu_2,\bm{\kappa},\pm1/2,\pm1/2} \right|^2\right) \nonumber \\
  &\times \sum_{\mu_j=-j}^{j}
  \Large | \left(l,\mu_j\mp1/2,1/2, \pm1/2|j\mu_j\right) \nonumber \\
  &\times \mathcal{F}_{nlj,\mu_j\mp1/2}(\mathcal{K}) \Large|^2.
  \label{eq:sigma_pm}
\end{align}
For the $p_{1/2}$ orbit, we have 
\begin{equation}
 \sigma_{\pm} \propto 
 \frac{1}{3} \left| \mathcal{F}_{0,1,\frac{1}{2},0} \right|^2  
 + \frac{2}{3} \left| \mathcal{F}_{0,1,\frac{1}{2},\mp1} 
 \right|^2   
\end{equation}
and
\begin{equation}\label{eq:p1/2}
 \sigma_+ - \sigma_- \propto 
 \frac{2}{3} 
\left\{
\left| \mathcal{F}_{0,1,\frac{1}{2},-1} \right|^2  
-
\left| \mathcal{F}_{0,1,\frac{1}{2},+1} \right|^2   
\right\}.
 \end{equation}
Here we omit the product of $F_{\rm kin}$ and the squared sum of $\tilde{t}$ as a proportionality coefficient. 
Note that their signs are both positive.
For the $p_{3/2}$ orbit, 
\begin{equation}\label{eq:p3/2}
 \sigma_+ - \sigma_- \propto 
 \frac{2}{3} 
 \left\{ 
 \left| \mathcal{F}_{01\frac{3}{2},+1} \right|^2
 - \left| \mathcal{F}_{01\frac{3}{2},-1} \right|^2
 \right\}.
\end{equation}
Comparing Eqs.~\eqref{eq:p1/2} and \eqref{eq:p3/2}, one can find that the positive (negative) $m$ component gives positive (negative) contribution to $A_z \propto \sigma_+ - \sigma_-$, in $j_>$ ($j_<$) case.

\textit{Result and discussion.}
Now, we apply this theoretical framework to a calculation for the $^{16}$O$(\vec{p},2p)^{15}$N reaction at 250~MeV.
In the present calculation, kinematics are fixed as follows.
The particle-1 kinetic energy $T_1 = 158$~MeV, and its emission direction $\theta_1=27^{\circ}$, $\phi_1=0^{\circ}$ and particle-2 polar angle $\theta_2=56^{\circ}$.
Note that $\phi_1$ can be fixed at $0^\circ$ without loss of generality.
The particle-2 azimuthal angle $\phi_2$ is varied between $\phi_2=90^{\circ}$ and $270^{\circ}$ and the other degree of freedom is given by the conservation law.
The triple differential cross section (TDX) of the reaction is then given by $d^3\sigma/dT_1 d\Omega_1 d\Omega_2$.
The deviation of $\phi_{12} \equiv \phi_2 - \phi_1 $ from $180^\circ$ corresponds to the non-coplanarity, and TDX and $A_z$ are discussed as a function of $\phi_{12}$ in the present study.

The calculation is performed by using the computer program {\sc pikoe}~\cite{OGATA2024}.
As numerical inputs, we use the s.p. potential of Bohr and Mottelson~\cite{BOHR_MOTTELSON1969} for the nucleon s.p. wave function. 
The parametrization of Franey and Love~\cite{FRANEY_LOVE1985} is used for the $NN$ effective interaction, $t_{NN}$. 
The distorted waves of the incident proton and the emitted protons are obtained as scattering waves under the optical potentials of the EDAD1 parameter of the Dirac phenomenology~\cite{Hama90,Coo93}.
Note that the calculation is done using {\sc pikoe} code~\cite{OGATA2024} but a new option to calculate $A_z$ is implemented for the present study.
To realize calculations corresponding to Eq.~\eqref{eq:sigma_pm}, the LS term of the distorting potential is excluded and $C_{zz} = 1$ is explicitly assumed by imposing $\mu_0 = \mu_N$ in the numerical calculation, unless otherwise mentioned.

\begin{figure}[tbp]
    \centering
    \includegraphics[width=0.9\linewidth]{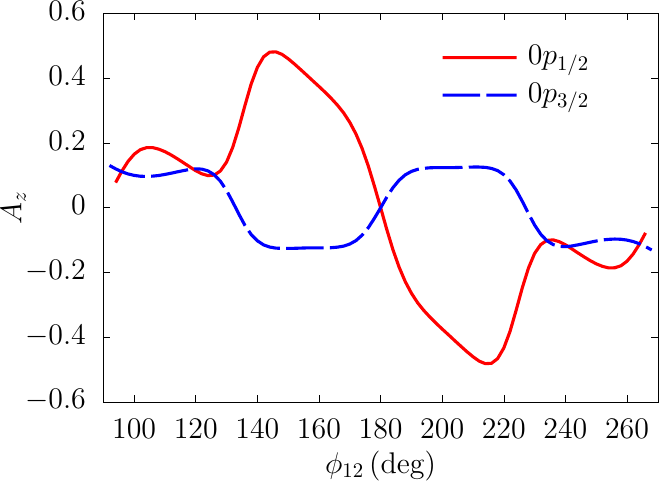}
    \caption{
      $A_z$ of $^{16}$O$(\vec{p},2p)$ knockout reaction from $0p$ orbits calculated using Eq.~\eqref{eq:sigma_pm}.
    }
    \label{fig:result_nols}
\end{figure}

In Fig.~\ref{fig:result_nols}, the $\phi_{12}$ dependence of calculated $A_z$ of $^{16}$O$(\vec{p},2p)$ knockout reaction from $0p$ orbits is shown.
It is confirmed that $A_z=0$ at $\phi_{12}=180^{\circ}$ (coplanar condition). 
In addition, $A_z$ has symmetry as $A_z(\phi_{12})=-A_z(360^{\circ}-\phi_{12})$. 
Note that the operation $\phi_{12}\rightarrow 360^{\circ}-\phi_{12}$ corresponds to the mirror reflection (i.e. $\mathcal{K}\rightarrow \mathcal{\tilde{K}})$, because the particles $1$ and $2$ are identical.
The region $\phi_{12}<180^{\circ}$ corresponds to the situation shown in Fig.~\ref{fig:reaction}.
In this case, $A_z$ has a positive value for the $0p_{1/2}$ orbit, as expected.
Also, as expected from Eqs.~(\ref{eq:p1/2}) and (\ref{eq:p3/2}), $A_z$ has the opposite sign between the $0p_{1/2}$ and $0p_{3/2}$ orbits. 
It is confirmed that the overall downward trend to the right originates from the LS term of the $NN$ effective interaction, by an analysis excluding that term.

\begin{figure}[tbp]
    \centering
    \includegraphics[width=0.9\linewidth]{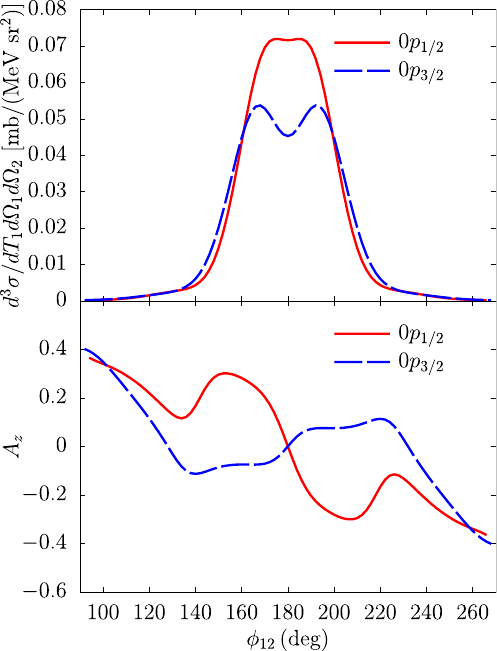}
    \caption{
      TDXs (top) and longitudinal analyzing powers (bottom) $A_z$ calculated for the $^{16}$O$(\vec{p},2p)$ reaction at 250~MeV as a function of $\phi_{12}$, including the LS term of the optical potential and the $\mu_0 \ne \mu_N$ components.
    The proton knockout from the $0p_{1/2}$ and $0p_{3/2}$ orbits are labeled as solid and dashed lines, respectively. 
    }
    \label{fig:results}
\end{figure}

More realistic calculations can be done by restoring the LS term of the optical potential and the $\mu_0 \ne \mu_N$ components.
In Fig.~\ref{fig:results}, TDX and $A_z$ of such calculations are shown.
In comparison with Fig.~\ref{fig:result_nols}, the behavior of $A_z$ in $140^\circ \lesssim \phi_{12} \lesssim 220^\circ$ is kept similar, and the aforementioned mechanism survives in the realistic case.
It is notable that $A_z$ has peaks at around $\phi_{12} = 140^\circ$--$150^\circ$ and $210^\circ$--$220^\circ$, where TDX has finite magnitude and promising for experimental measurements.
The LS term of the distorting potential has a major effect at both edges of the $\phi_{12}$ region, although TDX in this region is very small.

As a future prospect, $A_z$ can be defined for other knockout reactions at intermediate energies induced by the longitudinally polarized protons, if there are more than two particles in their final states.
In practice, however, a large $A_z$ value requires a strong spin-spin correlation between the incident proton and the particle to be knocked out. 
Therefore, we consider that it is difficult to apply this method to knockout reactions such as $(p,p\alpha)$ but may be used for studying $(p,p^3\mathrm{He})$ or $(p,pt)$ reactions, from which one could study the orbital motion of the $^3\mathrm{He}$ and $t$ clusters.

In the present study, we discussed $A_z$ in analogy with the Maris effect and its mechanism based on the absorption effect, selection of $\bm{K}_N$, and effective $NN$ interaction. 
As mentioned in the introduction, in the $\vec{d}\vec{p}$ breakup reaction study~\cite{Meyer04}, relatively small but similar behavior of $A_z$ is reported from both experimental data and theoretical calculations using realistic nuclear forces.
Up to here, we have shown the application of $A_z$ as a measure of axial vector component embedded in nuclear structures (i.e. the direction of $l$, when pinned by the spin orientation).  
On the other hand, we also consider that $A_z$ could be also sensitive to the axial vector coupling of nuclear interactions. 
For example, the rank-1 part of $3N$ force~\cite{FUKUI2024138839} might be investigated by $A_z$, as discussed in Ref.~\cite{Knutson94} for three-nucleon systems.
A possible advantage of this type of reaction is that the interacting nucleons are detected as scattering particles in the final state.
By suitably choosing the momenta and spin configurations of the two particles, one may selectively access the contributions from the terms of interest and suppress contributions from other terms.
A unique characteristic of $A_z$ is that it vanishes in the coplanar kinematical conditions, as discussed in the present study.
The dependence of $A_z$ on the deviation from the coplanar condition may provide important information.
Comprehensive measurements of spin observables, including $A_z$, should also be important.
In particular, a combined analysis using both $A_y$ and $A_z$ will be important for identifying the underlying origins of the observed asymmetry.
A discussion of suitable kinematical setups and spin observables for pinning down a specific term is in progress. 
From an experimental point of view, it is possible to measure $A_z$ by using a $\phi$-symmetric setup such as a solenoid spectrometer. This is in contrast to the case of $A_y$, where only a part of $\phi$ angles can contribute to $A_y$. Such a unique feature of $A_z$ is most suitable for studying the s.p. nature in unstable nuclei using radioactive isotope beams with the $(p,pN)$ reaction in inverse kinematics~\cite{KAWASE2018}.   

\textit{Summary.}
We proposed to use a longitudinally polarized proton to induce chirality in the $(p,pN)$ reaction at intermediate energies. 
Using the non-coplanar three-body kinematics in the final state, the analyzing power $A_z$ was defined to distinguish the direction of the orbital motion for an s.p. wave function, whose spin direction is effectively polarized along the beam direction due to the strong spin correlation with the incident proton ($C_{zz}\approx 1$). 
We investigated this effect for the $^{16}$O($\vec{p},2p)$ reaction at 250~MeV based on the intuitive picture, as well as the formulation based on the DWIA for the realistic calculations. The result shows large $A_z$ values for the $0p_{1/2}$ and $0p_{3/2}$ orbits, exhibiting a characteristic behavior; the sign of $A_z$ for an orbit is opposite to its LS partner. It can be regarded that the chirality induced by the incident proton is delivered to the chirality of the final states through s.p. states with the LS coupling and the absorption effect of the nucleus.
 
\begin{acknowledgments}
We wish to acknowledge 
T. Uesaka
for valuable comments and discussions.
This work is supported in part by Grant-in-Aid for Scientific Research (Nos.\ JP20K14475, JP21H04975, JP25K07302, and JP25K17400) from Japan Society for the Promotion of Science (JSPS), and JST ERATO Grant No. JPMJER2304, Japan. 
\end{acknowledgments}

\bibliography{az.bib}


\end{document}